\documentstyle[epsf,12pt]{article}
\begin{document}
\begin{center}
\Large{\bf A Case for the Chiral Magnetic Effect in mid-peripheral Au-Au 
Collisions $\sqrt{s_{NN}}$=200 GeV.}\\
\large{R.S. Longacre$^a$\\
$^a$Brookhaven National Laboratory, Upton, NY 11973, USA}
\end{center}
 
\begin{abstract}
The Chiral Magnetic Effect (CME) is predicted for mid-peripheral Au-Au 
collisions $\sqrt{s_{NN}}$=200 GeV at RHIC. However many backgrounds can give 
signals that make the measurement hard to interpret. The STAR experiment has 
made a measurement that makes it possible to separate out the background from 
the signal. An event shape analysis is the key for controlling the effects of 
Transverse Momentum Conservation (TMC) in the Au-Au collisions. LHC Pb-Pb 
collision $\sqrt{s_{NN}}$=2.76 TeV has made event plane measurements from
which we calculate the CME and make predictions about what the event shape
analysis should give.
\end{abstract}
 
\section{Introduction} 

Topological configurations should occur in the hot Quantum Chromodynamic (QCD) 
vacuum of the Quark-Gluon Plasma (QGP) which can be created in heavy ion 
collisions. These topological configurations form domains of local strong 
parity violation (P-odd domains) in the hot QCD matter through the so-called 
sphaleron transitions. The domains might be detected using the Chiral Magnetic
Effect (CME)\cite{warringa} where the strong external (electrodynamic) magnetic
field at the early stage of a (non-central) collision, through the sphaleron 
transitions induces a charge separation along the direction of the magnetic 
field which is perpendicular to the reaction plane. Such an out of plane charge
separation, however, varies its orientation from event to event, either 
parallel or anti-parallel to the magnetic field (sphaleron or antisphaleron).
Also the magnetic field can be up or down with respect to the reaction plane
depending if the ions pass in a clockwise or anti-clockwise manner. Any P-odd
observable will vanish and only the variance of such observable may be 
detected.

The STAR collaboration\cite{STAR} has published a measurement of charge 
particle azimuthal correlations consistent with CME expectations. In 
Ref.\cite{voloshin} and used by STAR the CME can be indirectly approached
through a two-particle azimuthal correlation given by
\begin{equation}
\gamma = \langle cos(\phi_1 + \phi_2 - 2\Psi_{RP}) \rangle = \langle cos(\phi_1-\Psi_{RP}) cos(\phi_2-\Psi_{RP}) \rangle - \langle sin(\phi_1-\Psi_{RP}) sin(\phi_2-\Psi_{RP}) \rangle ,
\end{equation}
where $\Psi_{RP}$, $\phi_1$, $\phi_2$ denote the azimuthal angles of the
reaction plane, produced particle 1, and produced particle 2. This two
particle azimuthal correlation measures the difference between the in plane
and out of plane projected azimuthal correlation. If we would rotate all events
such that $\Psi_{RP}$ = 0.0,
then $\gamma$ would become
\begin{equation}
\gamma = \langle cos(\phi_1 + \phi_2) \rangle = \langle cos(\phi_1) cos(\phi_2) \rangle - \langle sin(\phi_1) sin(\phi_2) \rangle  .
\end{equation}
The CME predicts that $\gamma$ $>$ 0 for opposite sign-pairs and $\gamma$ $<$ 0
for same sign-pairs. There are other two particle azimuthal correlation effects
that can depend on the reaction plane driven by elliptic flow even though the
underlying correlation may be independent of the reaction plane. These 
backgrounds are summarized in Ref.\cite{koch}. It was pointed in 
Ref.\cite{koch} that the $\phi$ difference correlation ($\delta$) which is 
independent of the reaction plane gives a constraint on the CME and 
backgrounds.
\begin{equation}
\delta = \langle cos(\phi_1 - \phi_2) \rangle = \langle cos(\phi_1) cos(\phi_2) \rangle + \langle sin(\phi_1) sin(\phi_2) \rangle  .
\end{equation}

In Ref.\cite{koch} Transverse Momentum Conservation (TMC) is derived and
demonstrated that if there is no other correlation in the data except
elliptic flow TMC will give a negative
$\langle cos(\phi_1 + \phi_2 - 2\Psi_{RP}) \rangle$ ($\gamma$) which is
$v_2$ smaller than $\langle cos(\phi_1 - \phi_2) \rangle$ ($\delta$).
$\delta$ is a negative number given by TMC and scales as 1/N (N is the number
of particles). Depending on other correlations the connection between $\gamma$
and $\delta$ could be larger than $v_2$. Also in Ref.\cite{koch} Local Charge
Conservation (LCC) is another important background and the details of LCC is
found in Ref.\cite{pratt}. The authors of Ref.\cite{pratt} point out that for
same sign pairs LCC should give a small negative sign, while for opposite sign
pairs LCC should give a large positive correlation. Using the same coupling
effect to the reaction plane as TMC, one should expect
$\langle cos(\phi_1 + \phi_2 - 2\Psi_{RP})_{+-} \rangle$ ($\gamma$) equal $v_2$
times $\langle cos(\phi_1 - \phi_2)_{+-} \rangle$ ($\delta$) for the LCC.

Ref.\cite{warringa} has pointed out that P-odd domains on the surface of the
fireball omit same charge sign particles in the direction of the magnetic 
field. The particles that escape the surface would be of the same sign while 
the charge particles moving in the opposite direction would be of opposite 
sign. These particles would run into the fireball and be thermalized and loss
their direction (quenched). This effect makes it hard to know what the CME
should give for opposite sign pairs. Also when one considers opposite 
sign pairs the LCC effect is much stronger than the TMC effect making $\delta$ 
a large positive number. When we scale down by the expected $v_2$ to obtain
$\gamma$ we obtain a larger positive number which is bigger than what is 
measured. $\delta$ (2.0) times $v_2$ (.06) equals $\gamma$ (.12) where the 
measured value of $\gamma$ is .06. We need some additional negative .06 from 
some other source. Minijet quenching can increase the $\langle sin(\phi_1) 
sin(\phi_2) \rangle$ term because there is more quenching out of plane than in
plane. This could account for a reduction of .06. I think it is clear that 
having three important effects plus two quenching effects makes the opposite 
sign very hard to make an analysis. On the other hand LCC, minijets and 
CME quenching for the same sign can be considered unimportant.

The paper is organized in the following manner:

Sec. 1 is the introduction to correlations. Sec. 2 is an analysis of the same
sign data. Sec. 3 LHC analysis and predictions. Sec. 4 presents the summary 
and discussion.

\section{Same Sign Analysis} 

In performing the same charge sign analysis we will assume that only TMC and
CME amplitudes are present in the Au-Au $\sqrt{s_{NN}}$=200 GeV data. We will
use $\langle cos(\phi_1 + \phi_2 - 2\Psi_{RP}) \rangle$ ($\gamma$) and
$\langle cos(\phi_1 - \phi_2) \rangle$ ($\delta$) of Ref.\cite{STAR} in our 
analysis. Figure 1 is the $\gamma$ measurement of STAR while Figure 2 is the 
$\delta$ measurement. We can separate the Transverse Momentum Conservation 
(TMC) effect in the Au-Au collisions using a STAR event shape analysis.
For 20-40\% centrality half of the tracks of the TPC are used in order to
determine the reaction plane. The other tracks are used to calculate
$\langle cos(\phi_1 + \phi_2 - 2\Psi_{RP}) \rangle$ ($\gamma$) for same and
opposite sign. Also $\langle cos2(\phi - \Psi_{RP}) \rangle$ ($v_2$) is
calculated for these tracks. The $v_2$ measures the shape of each subevent.
For positive $v_2$ we have more tracks in plane, while for negative $v_2$ we
will have more tracks out of plane. The value for TMC in the correlator
$\gamma$ should be zero when $v_2$ is zero. The value should be negative when
$v_2$ is positive and positive when $v_2$ is negative. On the other hand the
CME should not depend on the event shape ($v_2$) but only on the collision 
geometry which is given by the centrality. Thus the magnetic field and the 
charge separation is a constant offset. 

In Figure 3 the
results are plotted from the STAR analysis which used correlators that are
proportional to -$\gamma$. Ref.\cite{koch} used measurements from 
Ref.\cite{STAR} for centrality range 50\% to 60\% because this should give the
best measurement signal to errors. The value for  $\langle cos(\phi_1 + \phi_2 
- 2\Psi_{RP}) \rangle$ ($\gamma$) is -.00045 $\pm$ .00005 and $\langle 
cos(\phi_1 - \phi_2) \rangle$ ($\delta$) is -.00038 $\pm$ .00010. $v_2$ is
0.06 and from Figure 3 assuming that the data for the same charge sign will be 
proportional to the centrality range 50\% to 60\%, we can determine the ratio
of pure CME to $\gamma$ at measured $\langle cos2(\phi_-\Psi_{RP}) \rangle$ 
$=$ 0.0 and $\langle cos2(\phi_-\Psi_{RP}) \rangle$ $=$ .06. Figure 3 gives
a ratio of 0.60 $\pm$ 0.15.

\begin{figure}
\begin{center}
\mbox{
   \epsfysize 5.0in
   \epsfbox{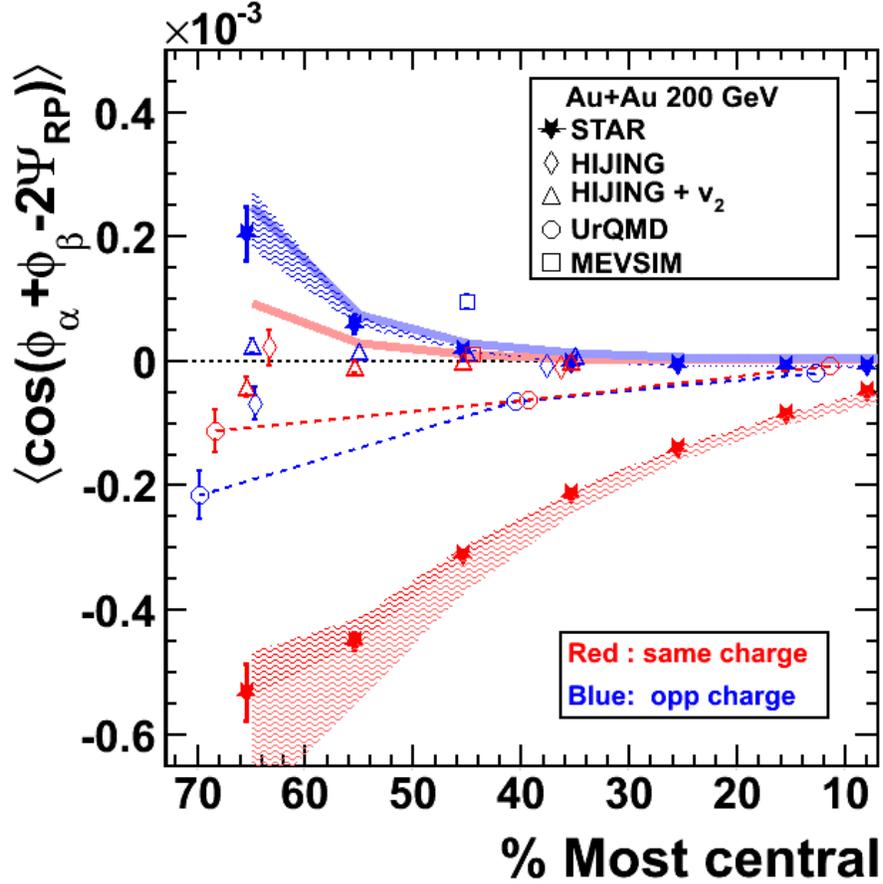}}
\end{center}
\vspace{2pt}
\caption{ (Color) (Solid points) $\langle cos(\phi_1 + \phi_2 - 2\Psi_{RP}) 
\rangle$ ($\gamma$) calculated for Au-Au $\sqrt{s_{NN}}$=200 GeV events data
acceptance cuts of 0.15 $<$ $p_t$ $<$ 2 GeV/c and $|\eta|$ $<$ 1.0. 
Event generators HIJING used the same input and acceptance cuts with and 
without an elliptic flow afterburner. Also generators URQMD and MEVSIM where 
used. Blue symbols mark opposite charge correlations, and red are same charge. 
For MEVSIM, HIJING, and URQMD points the true reaction plane from the
generated event was used for $\Psi_{RP}$. Thick, solid lighter-colored lines
represent possible non-reaction-plane-dependent contribution from many-particle
clusters, as estimated by HIJING and discussed in Sec. VIIA of Ref.\cite{STAR}.
}
\label{fig1}
\end{figure}

\begin{figure}
\begin{center}
\mbox{
   \epsfysize 5.0in
   \epsfbox{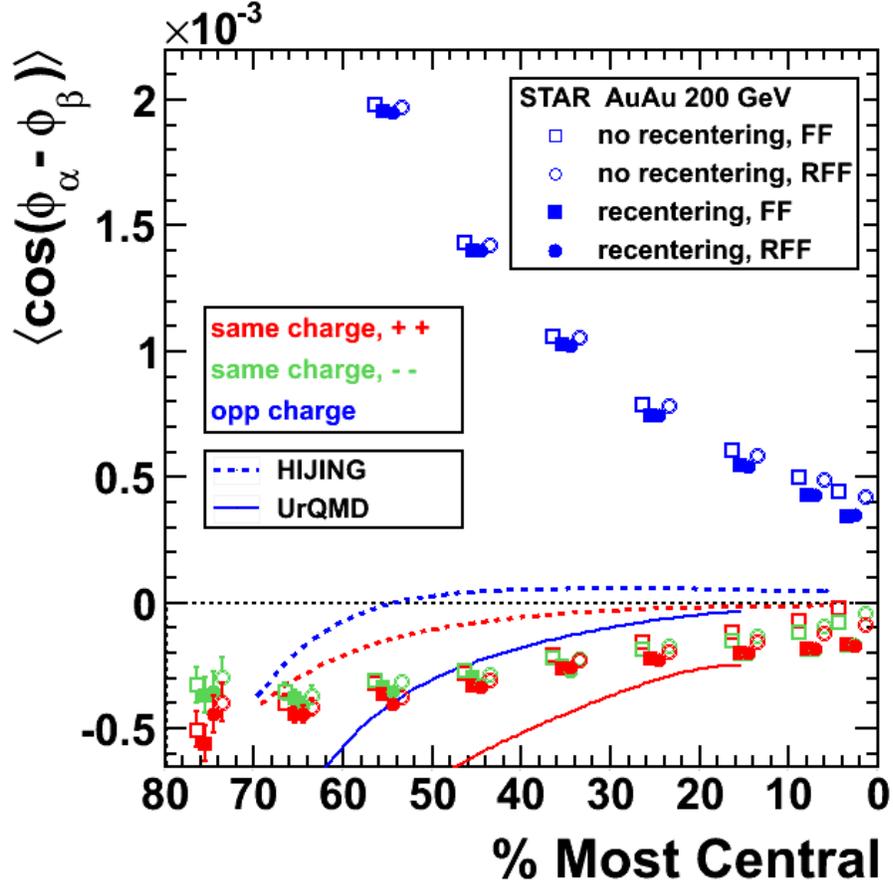}}
\end{center}
\vspace{2pt}
\caption{ (Color) $\langle cos(\phi_1 - \phi_2) \rangle$ ($\delta$) calculated 
for Au-Au $\sqrt{s_{NN}}$=200 GeV events data acceptance cuts of 0.15 $<$ 
$p_t$ $<$ 2 GeV/c and $|\eta|$ $<$ 1.0. We show different charge combinations 
and forward full magnetic field (FF) and reverse full magnetic field (RFF) 
configurations. The data points corresponding to different charge and field 
configurations are slightly shifted in the horizontal direction with respect 
to each other for clarity. The error bars are statistical. Also shown are 
model predictions described in Sec. VII of Ref.\cite{STAR}.}
\label{fig2}
\end{figure}
 
\begin{figure}
\begin{center}
\mbox{
   \epsfysize 5.0in
   \epsfbox{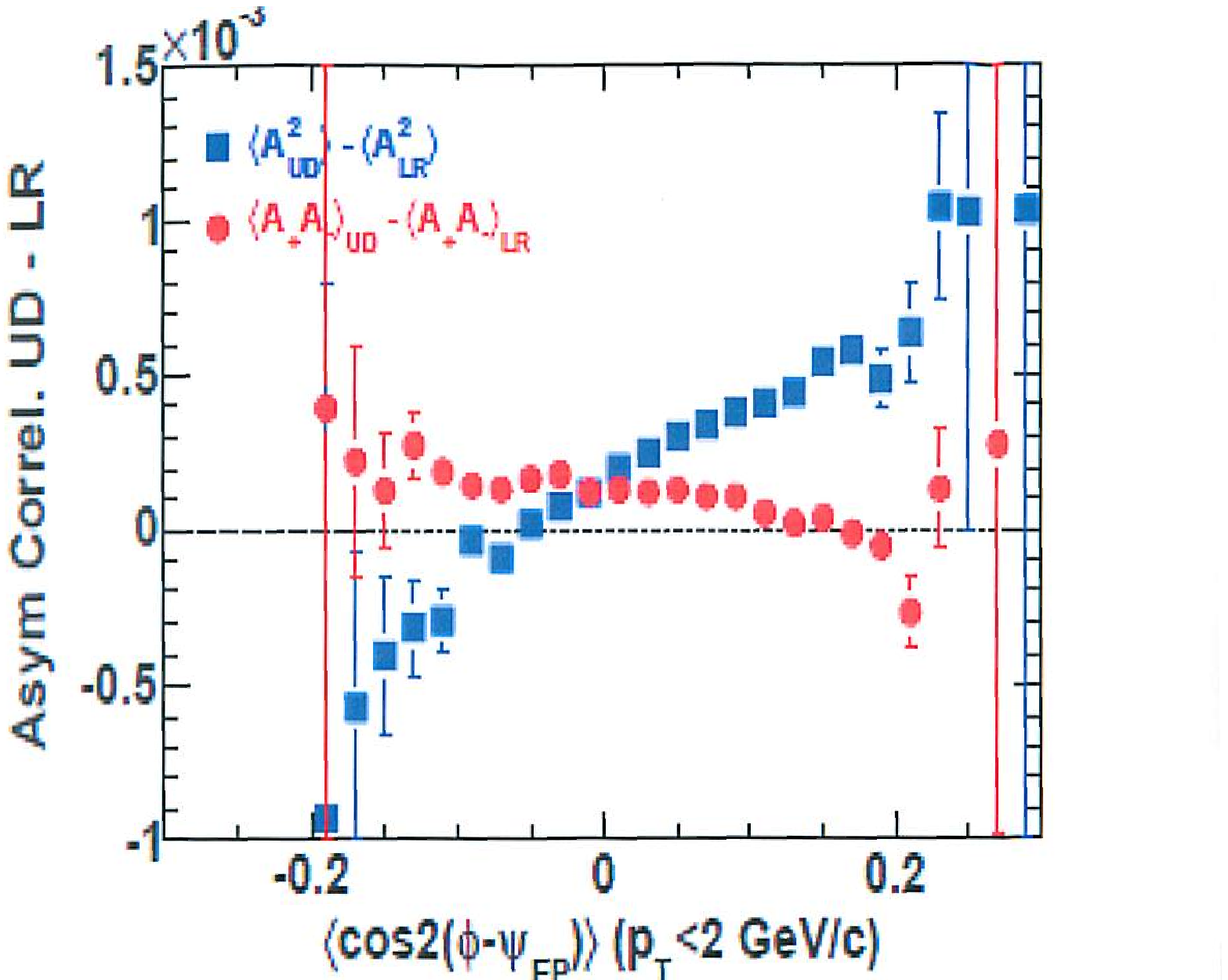}}
\end{center}
\vspace{2pt}
\caption{The charge correlation difference $\langle \rm A^2_{UD} \rangle - 
\langle A^2_{LR} \rangle$ and $\langle \rm A_+ A_- \rangle_{UD} - A_+ A_- 
\rangle_{LR}$ which is proportional to $-\gamma$ for the same sign pairs 
(squares) and opposite sign pairs (dots) under the selection of soft particles 
low-$p_t$ ($p_t$ $<$ 2 GeV/c) is plotted as a function of azimuthal anisotropy 
shape of the sub-events.}
\label{fig3}
\end{figure}

\subsection{Fit Model}

The fit model that we will use assumes that the three experimental values
$\gamma$, $\delta$ and the ratio for the centrality range 50\% to 60\% can be
fitted by three equations.
\begin{equation}
\gamma = \langle cos(\phi_1 + \phi_2) \rangle = \rm CME + ScaleT*TMC.
\end{equation}
\begin{equation}
\delta = \langle cos(\phi_1 - \phi_2) \rangle = \rm TMC - ScaleM*CME.
\end{equation}
\begin{equation}
 \rm ratio = {CME\over\gamma} =  {CME\over{CME + ScaleT*TMC}}.
\end{equation}
Where ScaleT is the faction of the TMC which is measure as part of $\delta$ as
translated to $\gamma$ and ScaleM is the faction of the CME which is measure as
part of $\gamma$ as translated to $\delta$. From the above equations we have
three equations and four unknowns.

\subsection{In Plane and Out of Plane}

Equation 2 and equation 3 show how $\gamma$ and $\delta$ can be broken up into
an in plane term $\langle cos(\phi_1) cos(\phi_2) \rangle$ and an out of plane 
term  $\langle sin(\phi_1) sin(\phi_2) \rangle$. If we would make a simple
assumption that the in plane $\langle cos(\phi_1) cos(\phi_2) \rangle$ is equal
to the TMC value and that the out of plane $\langle sin(\phi_1) sin(\phi_2) 
\rangle$ is equal to the CME value, one can fit the equation 4 and equation 5
with ScaleT and ScaleM equal 1.0. TMC is equal to -.00041 and the CME is equal
to -.00004, but the ratio is only .09 which is way off. Let us consider the
behavior of the CME term. Let us look at the pure CME case. Equation 4 becomes 
\begin{equation}
\langle cos(\phi_1) cos(\phi_2) \rangle - \langle sin(\phi_1) sin(\phi_2) \rangle  \rm = CME,
\end{equation}
while equation 5 becomes
\begin{equation}
\langle cos(\phi_1) cos(\phi_2) \rangle + \langle sin(\phi_1) sin(\phi_2) \rangle \rm = -ScaleM*CME .
\end{equation}

$\langle cos(\phi_1) cos(\phi_2) \rangle$ and  $\langle sin(\phi_1) sin(\phi_2)
\rangle$ can be written as
\begin{equation}
2\langle cos(\phi_1) cos(\phi_2) \rangle = \rm (CME - ScaleM*CME),
\end{equation}
and
\begin{equation}
2\langle sin(\phi_1) sin(\phi_2) \rangle = \rm -(CME + ScaleM*CME).
\end{equation}
We know that CME is left right symmetric therefore $\langle cos(\phi_1) 
cos(\phi_2) \rangle$ term equation is zero thus 
\begin{equation}
\rm 0.0 =  (CME - ScaleM*CME),
\end{equation}
or ScaleM = 1.0. This is true in general.
\subsection{Final Fit}

We have seen in the last subsection that in general ScaleM = 1.0 thus
equation 4 through 6 becomes
\begin{equation}
\gamma = \langle cos(\phi_1 + \phi_2) \rangle = \rm CME + ScaleT*TMC.
\end{equation}
\begin{equation}
\delta = \langle cos(\phi_1 - \phi_2) \rangle = \rm TMC - CME.
\end{equation}
\begin{equation}
\rm ratio = {CME\over\gamma} =  {CME\over{CME + ScaleT*TMC}}.
\end{equation}
We now have three equations and three unknowns. After doing the fit we
obtain Table I.

\bf Table I. \rm Fit parameters and errors.

\begin{center}
\begin{tabular}{|c|r|r|}\hline
\multicolumn{3}{|c|}{Table I}\\ \hline
parameters & value & errors \\ \hline
CME & -.0002625 & +.0000638 -.0000845 \\ \hline
TMC & -.0006460 & +.0001195 -.0001295 \\ \hline
ScaleT & .2836 & +.1601 -.1296 \\ \hline
\end{tabular}
\end{center}

Ref.\cite{koch} made a prediction that ScaleT should be equal to $v_2$ or
0.06. However there must be some other correlation of same sign charge pairs
which makes the effect four times bigger. Ref.\cite{ron} showed that there
is a suppression of near side same sign charge pairs created by the
strong chromo-electric fields inside the glasma flux tubes generated by the
color glass initial state of the Au-Au collisions. This suppression greatly 
increases the negative TMC effect between back to back particles of the same
charge sign.

\section{LHC Analysis and Predictions}

At the LHC the ALICE experiment has made measurements like STAR but on Pb-Pb
at $\sqrt{s_{NN}}$=2.76 TeV. We will again performing a same charge sign 
analysis using $\langle cos(\phi_1 + \phi_2 - 2\Psi_{RP}) \rangle$ ($\gamma$) 
and $\langle cos(\phi_1 - \phi_2) \rangle$ ($\delta$). Figure 4 is the 
$\gamma$ measurement of ALICE while Figure 5 is the $\delta$ measurement. 
Since LHC is at such a higher energy hard scattering or jets play a larger 
role. Let us initially assume that only jets and CME are present in the data.

\begin{figure}
\begin{center}
\mbox{
   \epsfysize 5.0in
   \epsfbox{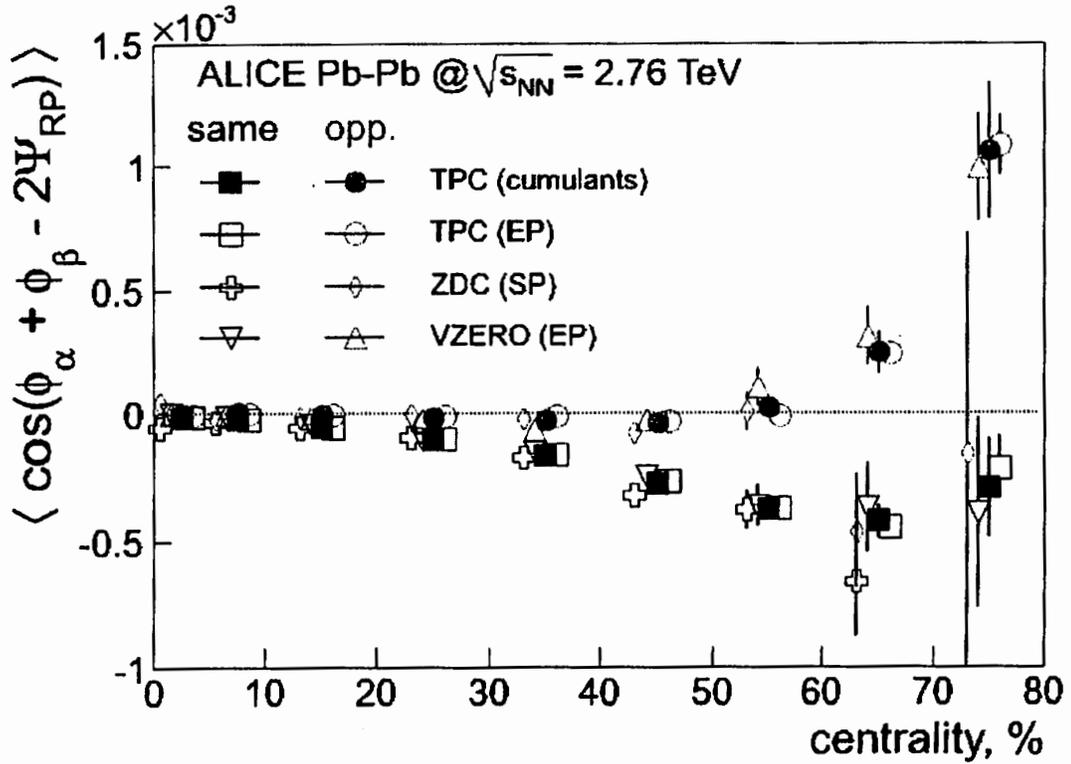}}
\end{center}
\vspace{2pt}
\caption{ The centrality dependence of the integrated three-particle correlator
$\langle cos(\phi_\alpha + \phi_\beta - 2\Psi_{RP}) \rangle$ ($\gamma$) 
measured with four independent methods: the cumulant method using TPC tracks; 
the event plane estimation using the TPC, using the ZDC, or using the VZERO 
detectors. The correlation between same charge pairs and opposite charge pairs 
are indicted on the figure.}
\label{fig4}
\end{figure}

\begin{figure}
\begin{center}
\mbox{
   \epsfysize 5.0in
   \epsfbox{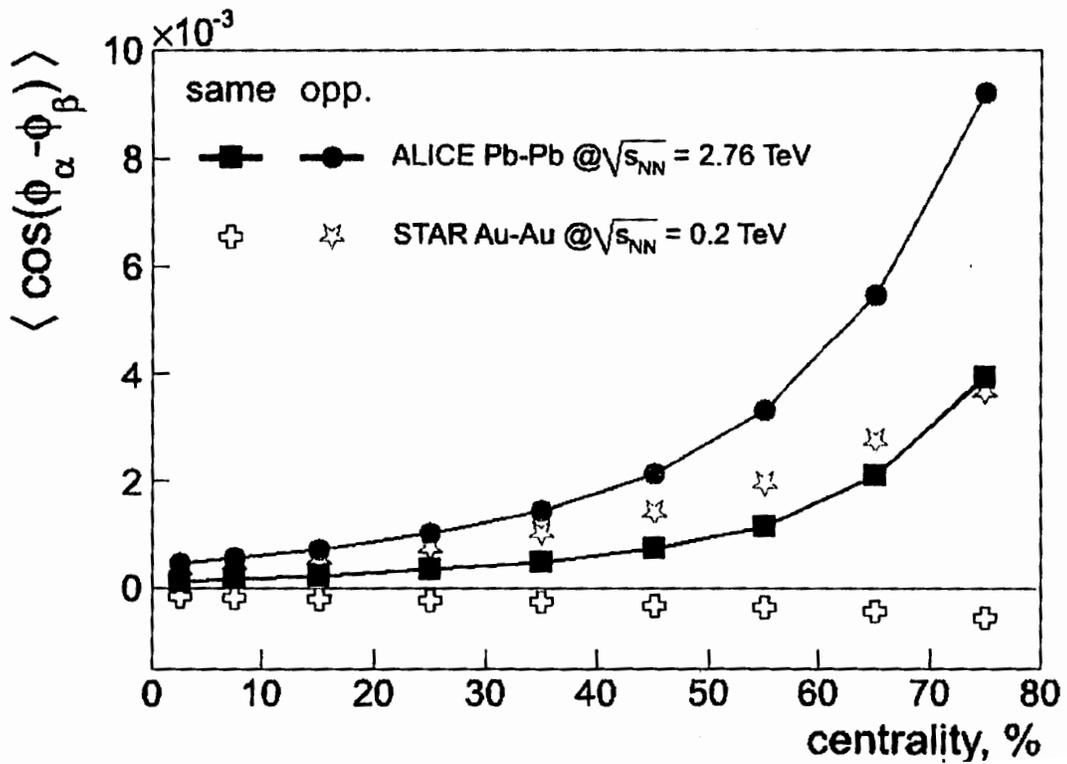}}
\end{center}
\vspace{2pt}
\caption{ The centrality dependence of the integrated two-particle correlator
$\langle cos(\phi_\alpha - \phi_\beta) \rangle$ ($\delta$) measured by ALICE
(close symbols) and STAR (open symbols). The correlation between same charge
pairs and opposite charge pairs are indicted on the figure.}
\label{fig5}
\end{figure}

Again the fit model will use the centrality range 50\% to 60\% for the LHC.
The value for  $\langle cos(\phi_1 + \phi_2 - 2\Psi_{RP}) \rangle$ ($\gamma$) 
is -.0004 and $\langle cos(\phi_1 - \phi_2) \rangle$ ($\delta$) is .00125.
The $v_2$ value for this data is .105. We break up the $\gamma$ and $\delta$ 
into an in plane term $\langle cos(\phi_1) cos(\phi_2) \rangle$ and an out of 
plane term  $\langle sin(\phi_1) sin(\phi_2) \rangle$. 
\begin{equation}
\langle cos(\phi_1) cos(\phi_2) \rangle = \rm .000425,
\end{equation}
and
\begin{equation}
\langle sin(\phi_1) sin(\phi_2) \rangle = \rm .000825.
\end{equation}

\subsection{Jets and CME}

Using Ref.\cite{koch} as applied to jets, we can write
\begin{equation}
\gamma = \langle cos(\phi_1 + \phi_2) \rangle = \rm v_2 Jet.
\end{equation}
\begin{equation}
\delta = \langle cos(\phi_1 - \phi_2) \rangle = \rm Jet.
\end{equation}
Thus solving for $\langle cos(\phi_1) cos(\phi_2) \rangle$, we have
\begin{equation}
\langle cos(\phi_1) cos(\phi_2) \rangle = \rm 0.5(1 + v_2) Jet = .000425,
\end{equation}
and
\begin{equation}
\langle sin(\phi_1) sin(\phi_2) \rangle = \rm 0.5(1 - v_2) Jet + CME = .000825.
\end{equation}
The Jet = .0007692 and CME = -.00048 which is bigger than the RHIC value
of -.00026.

\subsection{TMC at the LHC}

We need to add Transverse Momentum Conservation (TMC) to the LHC Pb-Pb same 
charge sign analysis. At first let us assume that scaling is not the same as 
for jets and could differ from $v_2$. We can rewrite equations 19 and 20 as
\begin{equation}
\langle cos(\phi_1) cos(\phi_2) \rangle = \rm 0.5(1 + v_2) Jet  - 0.5(1 + ScaleT) TMC = .000425,
\end{equation}
and
\begin{equation}
\langle sin(\phi_1) sin(\phi_2) \rangle = \rm 0.5(1 - v_2) Jet - 0.5(1 - ScaleT) TMC + CME = .000825.
\end{equation}
The TMC value can be written in terms of the Jet value using equation 21.
\begin{equation}
\rm TMC = {{0.5(1 + v_2) Jet  - .000425}\over{0.5(1 + ScaleT)}}.
\end{equation}

TMC is a negative correlation and in equation 21 and 22 this negative sign has
been taken into account making the TMC term of these equations positive. From 
equation 23 if TMC = 0.0 then Jet = .0007692. If we increase Jet then TMC will 
increase. How large is the Jet term in the LHC data? In the next subsection 
we use the opposite charge sign pairs in order to estimate the value of Jet 
and TMC.

\subsection{Opposite Sign Data and the TMC}

We can use the opposite charge sign data for centrality range 50\% to 60\% 
from Figure 4  and Figure 5. The value for  $\langle cos(\phi_1 + \phi_2) 
\rangle$ ($\gamma$) is 0.0 and $\langle cos(\phi_1 - \phi_2) \rangle$ 
($\delta$) is .0033. We break up the $\gamma$ and $\delta$ into an in plane 
term $\langle cos(\phi_1) cos(\phi_2) \rangle$ and an out of plane term  
$\langle sin(\phi_1) sin(\phi_2) \rangle$ and obtain
\begin{equation}
\langle cos(\phi_1) cos(\phi_2) \rangle = \rm (1 + v_2) Jet  - 0.5(1 + v_2) TMC = .00165.
\end{equation}
Note that for ScaleT we have used $v_2$ because for the opposite sign pairs
there is no local suppression of near side pairs which increases the negative 
TMC effect between back to back particles. Also the 0.5 factor in front of the
Jet term is gone. This doubling of the Jet term come from the fact that in
p-p the opposite sign correlation for jets is twice the same sign correlation.

\subsection{LHC Predictions}

In the last two subsections we have determined two equations and two unknowns,
namely the Jet term and the TMC term in equation 23 and equation 24. First 
let us consider the case where $v_2$ is .105 and ScaleT  = $v_2$. From these 
two equations one gets Jet = .002217 and TMC = .001448. When we substitute 
these values into equation 22 the CME value is -.00048. This value is the same 
as determined in the the first LHC subsection where there was no TMC term.

What would an analysis of event shape give at the LHC?  Like in the STAR event 
shape analysis we use half of the tracks of the TPC in order to determine the 
reaction plane. The other tracks are used to calculate $\langle 
cos(\phi_1 + \phi_2 - 2\Psi_{RP}) \rangle$ ($\gamma$) for same sign. 
Also $\langle cos2(\phi - \Psi_{RP}) \rangle$ ($v_2$) is also 
calculated for these same tracks. The $v_2$ measures the shape of each 
subevent. For positive $v_2$ we have more tracks in plane, while for negative 
$v_2$ we will have more tracks out of plane. The value for Jet plus TMC in the 
correlator -$\gamma$ should be zero when $v_2$ is zero. The value of Jet plus
TMC should be positive when $v_2$ is positive and negative when $v_2$ is 
negative. We have plotted Figure 3 again adding a solid line which is 
proportional to -$\gamma$ that would be measured at ALICE if ScaleT is equal
to $v_2$ (Figure 6). The slope of the solid line is negative instead of
positive like it is at RHIC. The difference in sign comes from the positive
value of $\langle cos(\phi_1 - \phi_2) \rangle$ ($\delta$) at LHC while at 
RHIC it is negative.

The effect of the TMC term in the $\gamma$ correlator was determined at RHIC
and was four times larger than $v_2$. The same suppression of near side same 
sign charge pairs which is created by the strong chromo-electric fields 
inside the glasma flux tubes generated by the color glass initial state of 
the Pb-Pb collisions maybe at work. This suppression could greatly 
increases the negative TMC effect between back to back particles of the same
charge sign. We do not know how strong this effect is at the LHC, but if we
use the same value of ScaleT measured at RHIC ScaleT = .2836, then
Jet = .002040 and TMC = .001094. When we substitute these values into 
equation 22 the CME value is -.000304. Again we can predict on Figure 6 
(this time a dashed line) what should be measured at ALICE if ScaleT is equal
to .2836. The slope of the dashed line has the same value as the solid line 
except it has the opposite sign. This slope is not as extreme as at RHIC. The 
slope is flatter than at RHIC because of a cancellation by the Jet term scaled 
down by $v_2$ (.105) with the TMC term scaled by ScaleT (.2836). At RHIC there 
is no Jet term.

\begin{figure}
\begin{center}
\mbox{
   \epsfysize 5.0in
   \epsfbox{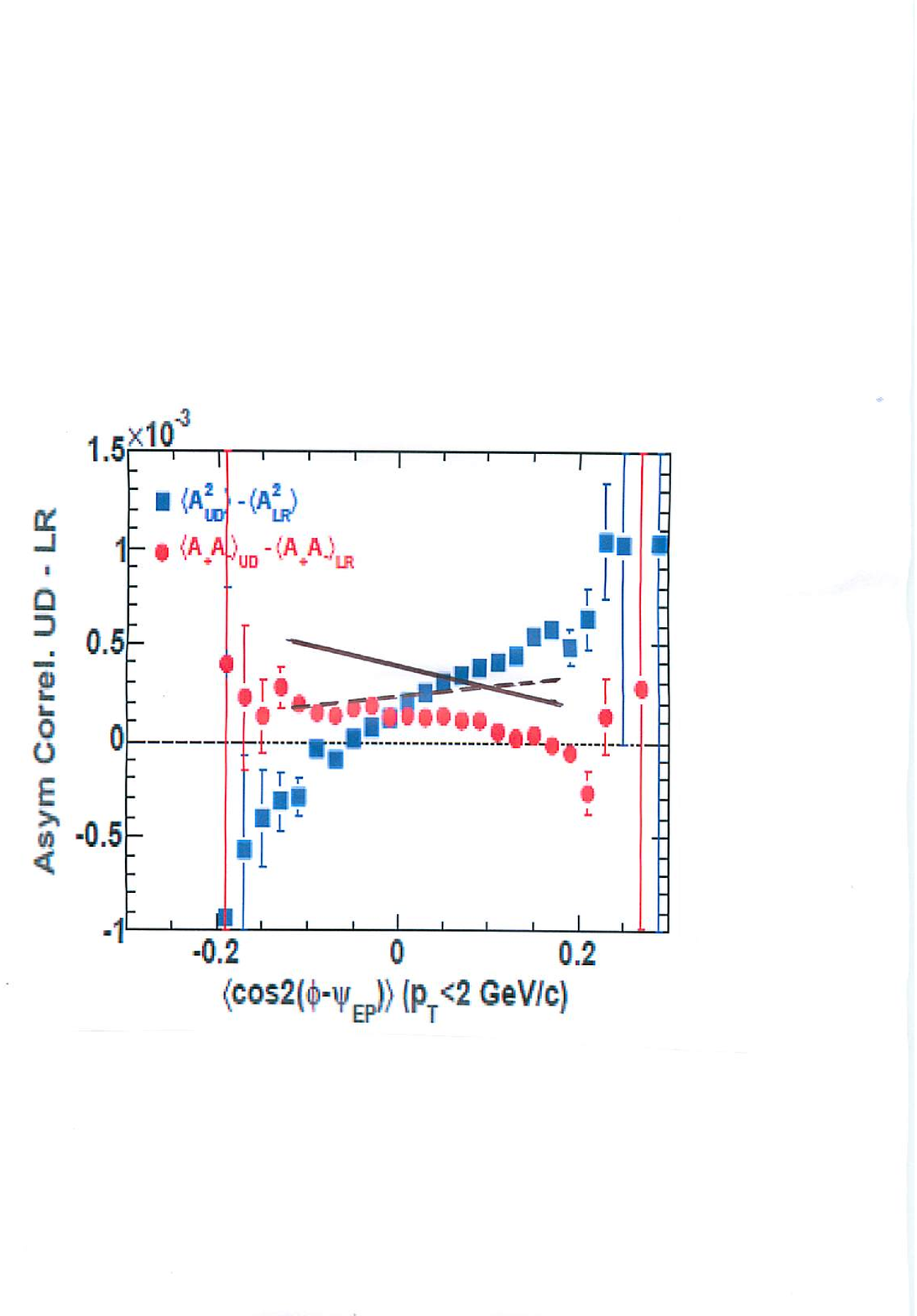}}
\end{center}
\vspace{2pt}
\caption{The STAR charge correlation difference $\langle \rm A^2_{UD} \rangle 
- \langle A^2_{LR} \rangle$ and $\langle \rm A_+ A_- \rangle_{UD} - A_+ A_- 
\rangle_{LR}$ which is proportional to $-\gamma$ for the same sign pairs 
(squares) and opposite sign pairs (dots) under the selection of soft particles 
low-$p_t$ ($p_t$ $<$ 2 GeV/c) is ploted as a function of azimuthal anisotropy 
shape of the sub-events. The solid line is a prediction of what would be 
measured at the LHC if the ScaleT is equal to $v_2$ (.105). The dash is a 
prediction of what would be measured at the LHC if the ScaleT is equal to 
what it was at RHIC (.2836).}
\label{fig6}
\end{figure}
 
\section{Summary and Discussion}

We have shown in this manuscript that The Chiral Magnetic Effect (CME) is
consistent with the correlation analysis for same sign charge pairs done by the
STAR experiment on mid-peripheral Au-Au Collisions $\sqrt{s_{NN}}$=200 GeV.
There are backgrounds present which can give signals that make the CME
measurement hard to interpret. The STAR experiment has made a measurement that 
makes it possible to separate out the background from the signal for the same 
sign charge pairs. For the same sign charge pairs the Transverse Momentum
Conservation (TMC) term is the only major background and it will very with the 
shape of particle production characterized by $\langle cos2(\phi - \Psi_{RP})
\rangle$ ($v_2$). The $v_2$ measures the shape of each event. For positive
$v_2$ we have more tracks in plane, while for negative $v_2$ we will have more
tracks out of plane. The value for TMC in the correlator
$\langle cos(\phi_1 + \phi_2 - 2\Psi_{RP}) \rangle$ ($\gamma$) should be zero
when $v_2$ is zero. The value should be negative when $v_2$ is positive and
positive when $v_2$ is negative. Thus through this event shape analysis one
can measure the CME.

We have made a fit to same charge pairs correlations at a centrality value
of 50-60\% using equations 12-14. We have determined that at this centrality
for Au-Au Collisions $\sqrt{s_{NN}}$=200 GeV the value of CME is -.0002625
+.0000638 -.0000845. The TMC is two and half times bigger and of the same
sign. The CME is a flow of charge out of the reaction driven by the magnetic 
field. This flow causes a charge separation and should be measurable. We can
define in an event by event manner an in plane and out of plane charge 
separation
\begin{equation}
\Delta \rm Q_{IN} = (N^L_+ - N^L_-) - (N^R_+ - N^R_-)
\end{equation}
and
\begin{equation}
\Delta \rm Q_{OUT} = (N^T_+ - N^T_-) - (N^B_+ - N^B_-).
\end{equation}
If there is a flow of charge out of plane then $\Delta \rm Q_{OUT}$ should be 
wider than the in plane $\Delta \rm Q_{IN}$. For centrality 40-50\% we show
the $\Delta \rm Q$ distributions (see Figure 7). In this Figure we also show 
that the rms difference between out of plane and in plane divided by the average
rms of the two distributions. The fact that this difference is positive and
not consistent with any known background other than CME, we conclude that the
CME is the best and simplest explanation of the data.    

\begin{figure}
\begin{center}
\mbox{
   \epsfysize 5.0in
   \epsfbox{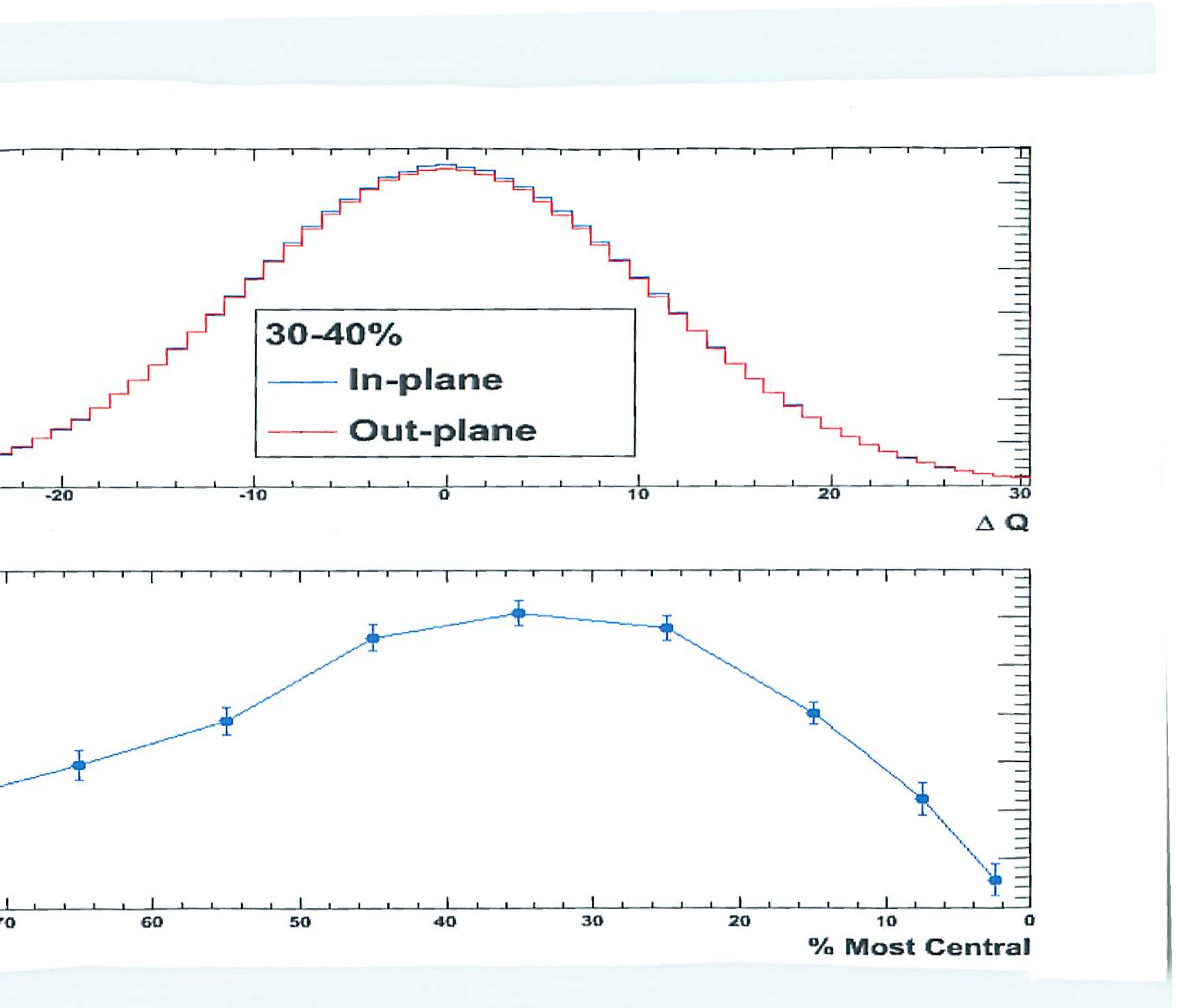}}
\end{center}
\vspace{2pt}
\caption{ Top panel: sample $\Delta Q$ distributions for 40-50\% Au-Au 
$\sqrt{s_{NN}}$ = 200 GeV collisions in plane and out of plane. Bottom panel:
is the rms difference between out of plane and in plane divided by the averge
rms of the two distributions.}    
\label{fig7}
\end{figure}

With the success of the RHIC Au-Au $\sqrt{s_{NN}}$=200 GeV CME measurement we
considered the LHC Pb-Pb $\sqrt{s_{NN}}$=2.76 TeV measurements. First we
find that $\langle cos(\phi_1 - \phi_2) \rangle$ ($\delta$) as measured by
ALICE for the same charge sign pairs has changed sign. We assume that this
is because of the importance of hard scattering or jets playing a greater
role. By adding a jet term we are able to determine the amount of CME for the
same 50-60\% centrality. The value of CME is a little less than twice the value
measured at RHIC -.00048 to -.00030. We can also make a prediction for the 
shape dependence of the event shape analysis if it is done at ALICE using 
their TPC (see Figure 6). This event shape analysis is a very important test of
our approach to the same charge sign correlation data. However if the TMC 
effects scales like it did at RHIC the sign of the event shape correlation 
slope sign will be the same as at RHIC but of a smaller value. If this 
prediction is false then our analysis is false. One should note the lower range
of the CME value falls with in the range of the RHIC result. One must also make
the $\Delta \rm Q$ measurements. The width of $\Delta \rm Q_{OUT}$ must 
continue to be larger than $\Delta \rm Q_{IN}$. 

\section{Acknowledgments}

This research was supported by the U.S. Department of Energy under Contract No.
DE-AC02-98CH10886.

\end{document}